%% file: manuscript.tex
\documentclass[pra,aps,showpacs, notitlepage, groupedaddress,superscriptaddress, twocolumn, numerical]{revtex4-1}
\input{./packages}

\newcommand{\er}[1]{\textcolor{black}{#1}}

\begin{document}

\title{Experimental verification of a reversed Clausius inequality in an isolated system}
\author{Daniel Mayer}
	\affiliation{Department of Physics and Research Center OPTIMAS, Technische Universit\"at Kaiserslautern, Germany}

	\author{Eric Lutz}
	\affiliation{Institute for Theoretical Physics I, University of Stuttgart, D-70550 Stuttgart, Germany}
	
	\author{Artur Widera}
	
	\affiliation{Department of Physics and Research Center OPTIMAS, Technische Universit\"at Kaiserslautern, Germany}

\begin{abstract}
	The second law of thermodynamics is a fundamental law of Nature. It is almost universally associated with the Clausius inequality that lower bounds a change in entropy  by the ratio of supplied heat and  temperature. However, this result presupposes that a system is in contact with  a heat bath that drives it to a thermal  state. For isolated systems that are moved from an initial equilibrium state  by a dissipative heat exchange, the Clausius inequality has been predicted to be reversed.
	We here experimentally investigate the nonequilibrium thermodynamics of an isolated  dilute gas  of ultracold Cesium atoms that can be either thermalized or pushed out of equilibrium by means of laser cooling techniques.  We determine in both cases the phase-space dynamics  by tracing the evolution of the gas with position-resolved fluorescence imaging, from which we evaluate all relevant thermodynamic quantities. 
	Our results confirm  the validity of the usual Clausius inequality for the first process and of the reversed Clausius inequality for the second transformation.
\end{abstract}
\maketitle

According to Clausius, the entropy variation  of a system during a thermodynamic process is greater than  or equal to the absorbed  heat divided by the  temperature at which that heat is absorbed,  $\Delta S \geq Q/T$ \cite{cla79}. This inequality is commonly regarded as a mathematical statement of the second law of thermodynamics \cite{pip66}. Its fundamental importance stems from the fact that  physical, chemical or biological transformations may be divided into three distinct categories depending on the sign of the inequality: nonequilibrium ($\Delta S > Q/T$), equilibrium  ($\Delta S = Q/T$) and impossible ($\Delta S < Q/T$) transformations. The Clausius inequality thus makes  a qualitative assertion about the  possible direction of a thermodynamic process and allows us to predict its evolution.

The theory of thermodynamics was originally developed for macroscopic systems (such as a gas contained in a cylinder) which inherently interact with an  environment  large  enough that it can be considered as a heat reservoir (the air around the cylinder, for instance) \cite{pip66}. The effect of such a heat bath is to thermalize any nonequilibrium state to an equilibrium Gibbs state. In the past decades, this framework has been successfully extended to nonequilibrium microscopic systems in contact with a heat reservoir \cite{bus05,sek10}. While the laws of thermodynamics still hold on average in this case, they have been generalized along single trajectories to account for nonnegligible thermal fluctuations \cite{sei12,jar11,cil13}. Stochastic thermodynamics has enabled the theoretical and experimental  study of the energetics of small systems, from colloidal particles to enzyme and molecular motors \cite{sei12,jar11,cil13}. 

At the same time, with the emergence of quantum technologies, many laboratories around the world have been investigating small systems that are highly isolated from their surroundings \cite{nie00,dow03}. Contrary to the assumptions of standard thermodynamics, nonequilibrium states do not necessarily thermalize in such a situation due to the lack of an external  heat bath \cite{eis15,gog16}.  A prominent example is provided by laser cooling  of atoms which plays an essential role in the study of new states of matter and high-resolution spectroscopy \cite{met99,coh11}.
Most laser cooling schemes  indeed only induce  thermalization of the momentum degrees of freedom \cite{met99,coh11}.
In dense atomic samples, frequent atomic collisions redistribute the energy and establish thermal equilibrium.
However,  in dilute  gases with  rare interparticle collisions, the absence of a heat reservoir leads  to far from equilibrium  states that do not thermalize on their own. The Clausius inequality has been shown to generally hold for nonequilibrium states that are driven by a heat reservoir to a  thermal state \cite{par89,per02,gav16}. By contrast, for initially isolated equilibrium states that dissipatively evolve into a nonthermal state, the Clausius inequality has been predicted to be   reversed,  $\Delta S \leq Q/T$ \cite{par89,per02,gav16}. While  the Clausius and reversed Clausius formulas agree for equilibrium processes, the two inequalities prognosticate exact opposite possible nonequilibrium transformations.

We here report the experimental study  of both inequalities using an isolated ultracold atomic sample of few noninteracting Cesium (Cs) atoms
in a crossed optical dipole trap \cite{met99,coh11}. Axial and radial trap directions are only weakly coupled, making the problem essentially one-dimensional. We prepare an equilibrium Gibbs state of the atomic system by applying a sufficiently long optical molasses pulse that thermalizes position and momentum degrees of freedom \cite{met99,coh11}.
We further employ pulses of degenerate Raman sideband cooling (DRSC) \cite{vul98,ker00,han00}  to drive the atomic sample out of equilibrium  by only thermalizing momentum coordinates. We measure the  position distribution of the atoms by means of position resolved in-situ fluorescence imaging \cite{sch16}. In combination  with numerical Monte-Carlo simulations of the three-dimensional trapping potential, we are able to determine the axial phase-space distribution of the system, from which we evaluate temperature, heat, entropy and entropy production. We confirm  the validity of the Clausius inequality for the thermalization  induced by an optical molasses pulse and present the first observation of the reversed Clausius inequality for the nonequilibrium transformation generated by  a degenerate Raman pulse.

\begin{figure*}[t]
	\begin{tikzpicture}
	\node (a) at (0,0) {\includegraphics[scale=1]{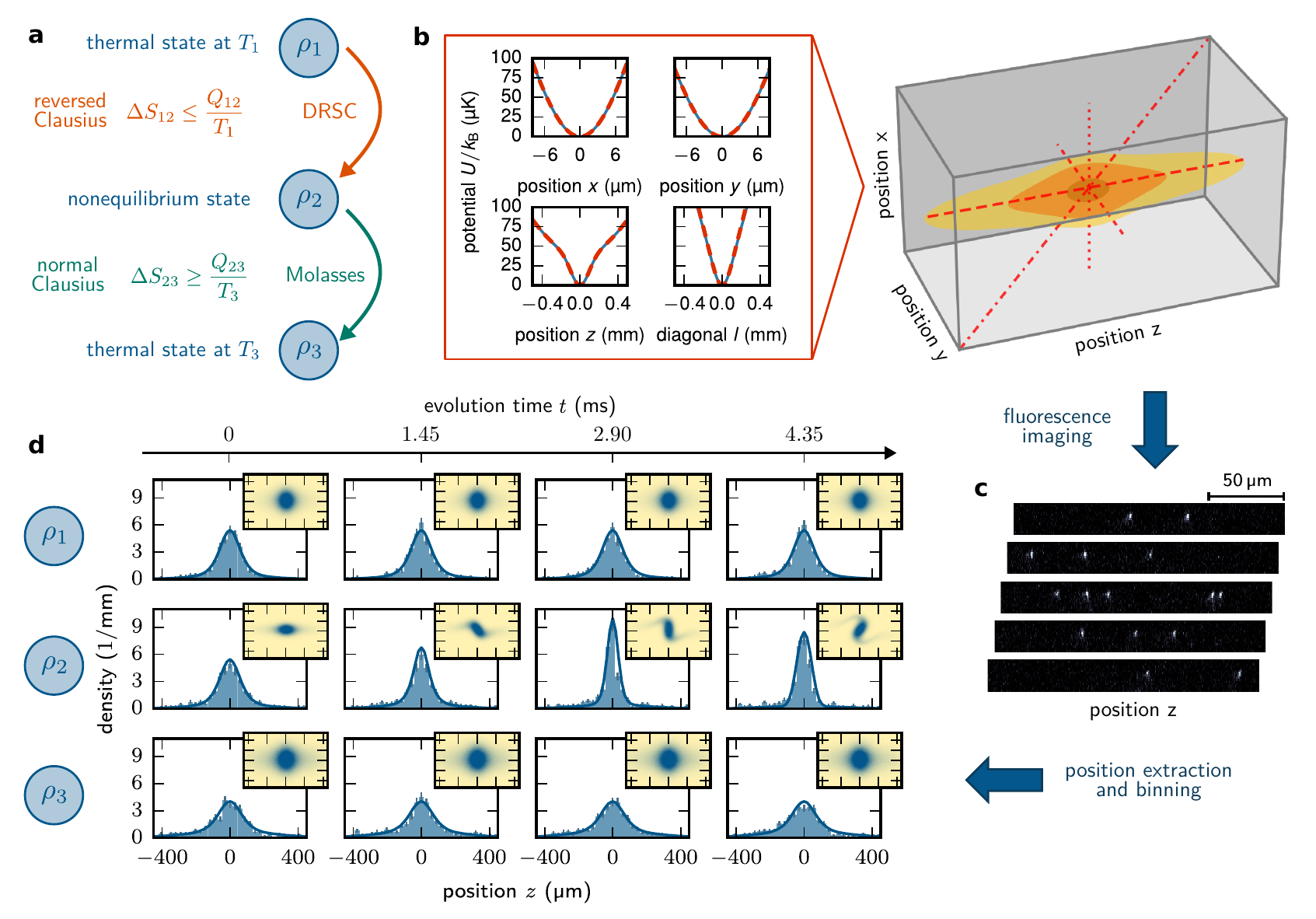}};
	\end{tikzpicture}
	\caption{
		\textbf{(a)}
		\er{Schematic of the two thermodynamic processes examined: i) transition from the equilibrium state $\rho_1$ to the nonequilibrium state $\rho_2$ induced by a degenerate Raman sideband cooling (DRSC) pulse and ii) transformation from the nonequilibrium state $\rho_2$ to the thermal state $\rho_3$ generated by an optical molasses pulse.}
		\textbf{(b)}
		Characterization of the dipole trap potential $U(x,y,z)$ used to store the Cesium (Cs) atoms.
		The red lines indicate the different directions of the potential cuts shown in the inset:
		\er{the full potential (blue lines) is there} compared to a separable potential approximation (red dashed lines) along the $x$-, $y$- and $z$-direction, as well as a diagonal cut  indicated by its length $l$.
		\textbf{(c)}
		Typical fluorescence images of the few Cs atom samples showing single Cs atoms as bright spots.
		By extracting the atomic positions along the $z$-axis from the images and binning them in a histogram, the atomic density distribution is measured.
		\textbf{(d)}
		For every state of the experimental protocol, the position distribution is measured at four evolution times, \er{$t= 0, 1.45, 2.90$ and $4.35$ ms} (blue bars).
		By combining the measured dynamics with a model (solid lines), the full axial phase-space distribution can be determined (insets).
 	}
	\label{fig:overview}
\end{figure*}

We begin by considering two thermodynamic processes (Fig.~1a): the first transformation, produced by a degenerate Raman pulse in the experiment,  brings an initial equilibrium state $\rho_1$ at temperature $T_1$ to a nonequilibrium state $\rho_2$, whereas the second transformation, triggered by an optical molasses pulse,  connects the nonequilibrium state $\rho_2$ to a  thermal Gibbs state $\rho_3$ at temperature $T_3$.  The variation of the entropy, $S_i= -k_\mathrm{B} \int dz dv_z \,\rho_i(z,v_z)$, where $\rho_i(z,v_z)$ is the (axial) phase-space density  of the system and $k_\mathrm{B}$ the Boltzmann constant, during the second  process satisfies \cite{pro76,sch80,esp10,def11},
\begin{equation}
\label{1}
	 \Delta S_\mathrm{23}  -\frac{Q_\mathrm{23}}{T_\mathrm{3}} = D(\rho_\mathrm{2}||\rho_\mathrm{3}) \geq 0,
\end{equation} 
where $Q_\mathrm{23}=  \int dzdv_z \,(\rho_3 - \rho_2) H$   is the heat absorbed by the system and  $H(z, v_z) = m_\mathrm{Cs} v_z^2/2 + U_z(z)$ its Hamiltonian. The  entropy production is given by the relative entropy between initial and final phase-space distributions,  $D(\rho_\mathrm{2}||\rho_\mathrm{3})=k_\mathrm{B}\int dzdv_z\, \rho_2\ln(\rho_2/\rho_3)$  \cite{cov06}. Since the latter quantity is nonnegative, Eq.~\eqref{1} implies the Clausius inequality. Formula \eqref{1} is thus a nonequilibrium extension of that inequality. On the other hand, the entropy difference during  the first transformation reads \cite{par89,per02,gav16},
\begin{equation}
\label{2}
	\Delta S_\mathrm{12} - \frac{Q_\mathrm{12}}{T_\mathrm{1}}  =- D(\rho_\mathrm{2}||\rho_\mathrm{1}) \leq 0.
\end{equation}
This is an expression of the reversed Clausius inequality that shows that the entropy change is upper bounded by the ratio of heat and temperature in this case. The physical difference between the two inequalities may be understood by noting that entropy is maximal at equilibrium \cite{jay57}. The entropy of a system thus  increases for a system that evolves from a nonequilibrium to an equilibrium state when in contact with a heat bath. By contrast, it decreases when a system is dissipatively driven away from equilibrium in the absence of a heat reservoir.

Ideal gases have played a seminal role in the study of statistical physics since Boltzmann \cite{bol64}. In our experiment, we initialize the system by trapping up to 12 Cs
atoms in a magneto-optical trap and then transfer them into a crossed
optical dipole trap. At typical temperatures of about $\SI{10}{\micro\kelvin}$, the atomic collision rate of $\SI{113}{\hertz}$ for the maximum number of 12 atoms is smaller than the inverse evolution time observed in the experiment. As a result, the system is effectively noninteracting. The crossed optical dipole trap at a wavelength $\lambda=\SI{1064}{\nano\meter}$ is formed by a horizontal trap beam (at a beam waist of $\SI{21}{\micro\meter}$ and power of $\SI{1}{\watt}$) and a second vertical beam (with waist of $\SI{165}{\micro\meter}$ and power of $\SI{8}{\watt}$).
Figure~\ref{fig:overview}b shows cuts through the total potential, $U(x, y, z)=U_x(x) + U_y(y)+ U_z(z)$, created by these Gaussian laser beams (including the contribution of gravity). The potential is strongly anharmonic but separable \er{(Fig.~1b)}. Atomic motions along radial and axial directions are hence decoupled (Supplemental Material), \er{and the problem is essentially one-dimensional}.

We first prepare the sample  in a thermal state $\rho_1$ at temperature $T_1$ by applying a (tunable) optical molasses onto the atoms (Supplemental Material).
By varying the detuning of the molasses cooling laser, the temperature $T_1$ can be controlled \er{between $\SI{16}{\micro\kelvin}$ and $\SI{21}{\micro\kelvin}$}.
We then drive the system to a nonequilibrium state $\rho_2$ with the help of a degenerate Raman pulse (Supplemental Material). To that end,  the atoms in the crossed dipole trap are pinned during the cooling pulse by a three dimensional optical lattice created by the DRSC laser beams, leaving their position distribution unchanged.
At the same time, the cooling effect of the DRSC redistributes the atomic velocities, leading to a Maxwell distribution of the (axial) velocity component
$ \tilde f_M(v_z) \propto \exp \left( {-m_\mathrm{Cs} v_z^2} / {2 k_\mathrm{B} T_\text{R}} \right) $
characterized by an effective Raman temperature $T_\text{R}$.
Here, $ \tilde f_i(v_z,t)$ generically denotes the momentum projection of the phase-space distribution $\rho_i(z,v_z,t)$.
We finally thermalize the atomic cloud  to a Gibbs state $\rho_3$ at temperature $T_3$  by applying a second pulse of optical molasses light.

We next measure the time evolution of these states  by extracting the (axial) $z$-positions of every single Cs atom from fluorescence images (Fig.~1c).  For that purpose, a one-dimensional optical lattice is  superimposed to the crossed dipole trap potential along the $z$-direction in order to freeze the atomic density distribution after a given  time $t$.
This allows us to access the time-dependent distributions $f_i(z, t)$,  the position projection of the phase-space density $\rho_i(z,v_z,t)$, for the three states as shown in Fig.~\ref{fig:overview}d.
The position distributions for the two equilibrium  states $\rho_1$ and $\rho_3$ do not vary in time, as expected. By contrast, the density-evolution of the nonequilibrium state $\rho_2$ features a breathing dynamics, visible as a contraction of the distribution  induced by the Raman pulse and the free phase-space evolution of the atomic cloud.

\begin{figure}[h]
	\begin{tikzpicture}
	\node (a) at (0,0) {\includegraphics[scale=1]{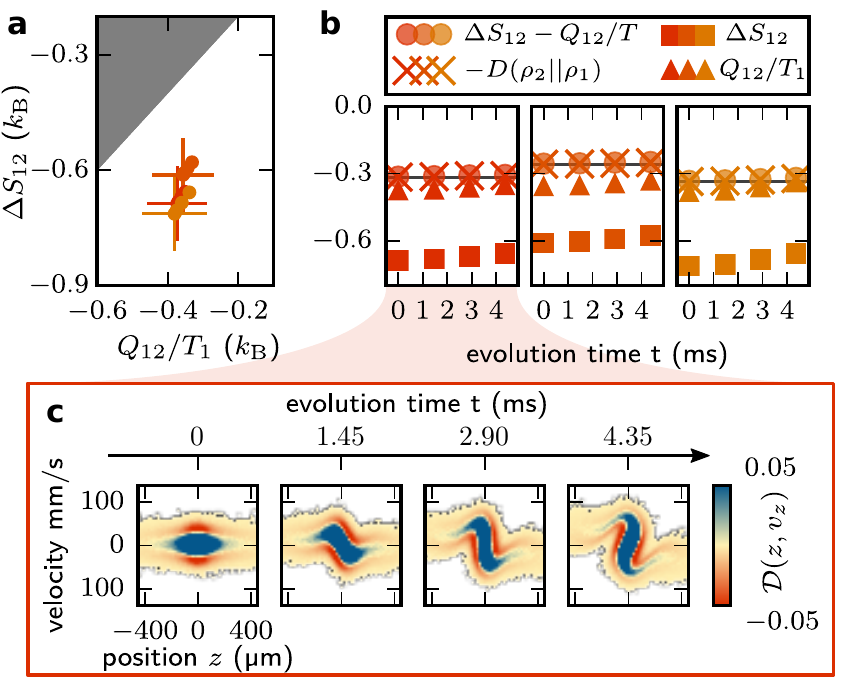}};	\node (a) at (0,-7.5) {\includegraphics[scale=1]{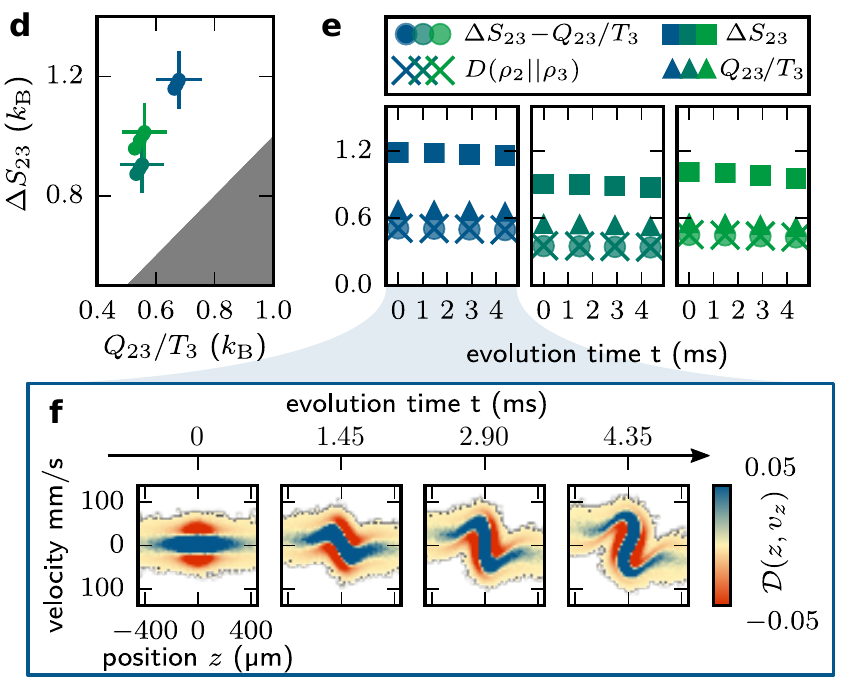}};
	\end{tikzpicture}
	\caption{
	\textbf{(a)}	
	\er{Verification of} the reversed Clausius inequality (2) for the transition from $\rho_1$ to $\rho_2$.
	The red, orange and yellow points correspond to molasses detunings of $\Delta f_\mathrm{cool} = \SI{-82}{\mega\hertz}$,  $\SI{-62}{\mega\hertz}$ and $\SI{-42}{\mega\hertz}$, respectively.
	\textbf{(b)}	
	\er{Time evolution of all thermodynamic} quantities in the reversed Clausius equality; \er{the solid gray line indicates the analytical result (\ref{3}). The small drift is due to the normalization of the phase-space density which is truncated to the experimentally accessible position range. }
	\textbf{(c)}	
	\er{Complex whorl dynamics emerging in the phase-space integrand $\mathcal{D}=\rho_1 \ln \left( \rho_1 / \rho_2 \right)$ of the relative entropy $D$ for  $\Delta f_\mathrm{cool} = \SI{-82}{\mega\hertz}$.
	\textbf{(d-f)}
	Analogous results to (a-c) for the  Clausius inequality (1).}	
	}
	\label{fig:clausius_results}
\end{figure}

We determine the temperature $T_1$ of the thermal state $\rho_1$  from the measured position distributions by comparing them to the expected (axial) Boltzmann  distribution
$ f_B(z) \propto \exp \left( { -U_z(z)}/{k_\mathrm{B} T_1} \right)$
for various values of the parameter $T_1$ in  a $\chi^2$-analysis.
Averaging over the temperatures  from all four measured times, $t= 0, 1.45, 2.90$ and $4.35$ ms, we find $T_1 = \SI{15.6\pm3.8}{\micro\kelvin}$. We similarly obtain  $T_3 = \SI{22.7\pm6.0}{\micro\kelvin}$ for the thermal state $\rho_3$.

The characterization of the nonequilibrium state $\rho_2$ proceeds as follows.
Right after the DRSC pulse at $t=0$, the velocity distribution is a Maxwellian at temperature $T_\text{R}$, while the position distribution is still a Boltzmannian at temperature  $T_1$.
Since $T_\text{R} \ne T_1$, both $f(z,t)$ and $\tilde{f}(v_z,t)$ will evolve in time. Noting that the Raman temperature  is the only free parameter of this nonequilibrium dynamics, $T_\text{R}$ can be extracted from the measured  position distributions by again employing a $\chi^2$-analysis as before: we get $T_\text{R} = \SI{4.4\pm3.3}{\micro\kelvin}$. A further consequence of the randomization effect of  a Raman pulse is that  the phase-space distribution factorizes, $\rho_2(z,v_z,t=0)= f_B(z) \tilde f_M(v_z)$, directly after such a pulse.  This factorization property also holds  for a thermal state. We can thus  determine the full axial phase-space distribution of all the three states,   immediately after each cooling pulse, from the subsequently measured evolution of the respective position distributions.

We now evaluate all the thermodynamic quantities needed to test the two Clausius inequalities \eqref{1} and \eqref{2}, such as heat, entropy and entropy production, from the phase-space densities $\rho_i$ $(i=1, 2, 3)$. Figure 2a shows the validation of the reversed Clausius inequality (2) for the first thermodynamic transformation that links the equilibrium state $\rho_1$ to the nonequilibrium state $\rho_2$ for three different values of the temperature $T_1$, corresponding to three laser detunings (red, orange and yellow points); the inaccessible region is indicated by the grey area. We emphasize that, according to  the standard Clausius inequality (1), these three processes are impossible, and hence not experimentally observable.  Figure 2b exhibits  both sides of the reversed Clausius equality (2), as a function of time, for the three temperature values. The equality, including the nonequilibrium entropy production (red, orange and yellow crosses), is found to be obeyed for all the measurement points. Remarkably, all the thermodynamic quantities are independent of time, even though the nonlinear microscopic dynamics, as represented by the integrand of the relative entropy, $\mathcal{D}(\rho_1,\rho_2) = \rho_1 \ln \left( \rho_1 / \rho_2 \right)= \mathcal{D}(z,v_z)$, exhibit complex whorl structures  (Fig.~2c) induced by the anharmonicity of the trap \cite{mil86}. This follows from the (quasi) Liouvillian evolution of the isolated atomic system. 

Since the position marginals $f_i(z)$ before and after the Raman pulse are the same, we may additionally use the additivity property of the relative entropy for independent distributions \cite{cov06} to derive an analytical expression for the nonequilibrium entropy production at $t=0$,
\begin{equation}
\label{3}
	D(\rho_2(0) || \rho_1) \!= \! D(\tilde{f}_2(0) || \tilde{f}_1) 
	\!=\! \frac{k_\mathrm{B}}{2} \!\left[ \ln\left(\frac{T_1}{T_\text{R}}\right) + \frac{T_\text{R}}{T_1} -1 \right].
\end{equation}
Equation \eqref{3} follows from the Gaussian form of the Maxwell velocity distribution \cite{cov06} and only depends on the initial temperature $T_1$ and the Raman temperature $T_\text{R}$. This analytical value is indicated by the horizontal line in Fig.~\ref{fig:clausius_results}b and shows excellent agreement.

\begin{figure}[t]
	\begin{tikzpicture}
	\node (a) at (0,0) {\includegraphics[scale=1]{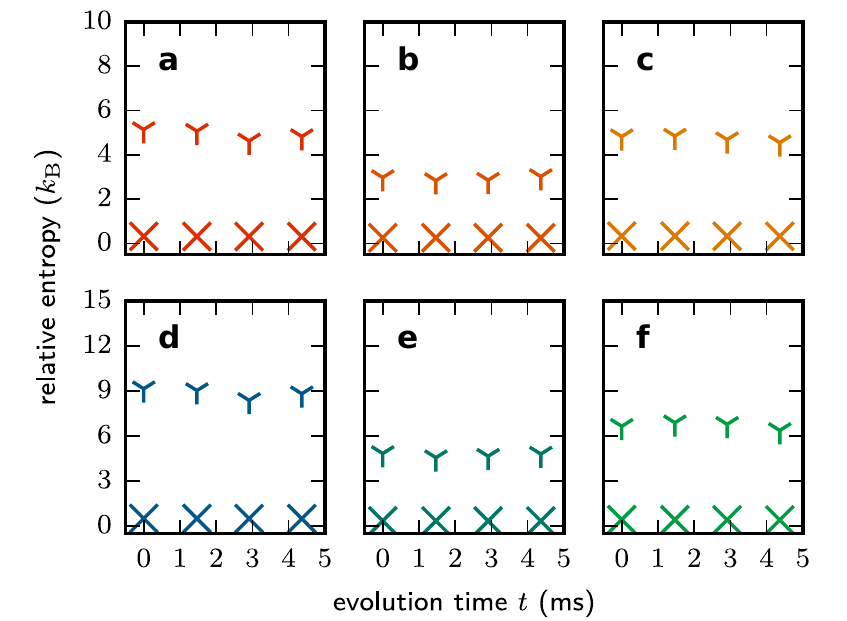}};
	\end{tikzpicture}
	\caption{\er{Relative entropy $D$ (crosses) and linear response approximation (4) (triangles) for the reversed \textbf{(a-c)} and the normal Clausius equality \textbf{(d-f)}.
		The  linear approximation deviates significantly from the exact value, indicating that the experiment is performed  far from the linear-response  regime.}}
	\label{fig:relent_approximation}
\end{figure}

We have repeated the same analysis for the second thermodynamic process that brings the nonequilibrium state $\rho_2$ to the thermal state $\rho_3$ (Figs.~2def), for three different temperatures $T_3$ (blue, teal and green points).  The Clausius inequality and  its nonequilibrium extension  \eqref{1} hold for this thermalization process, as expected.

We finally examine the linear response approximation of the entropy production, \er{which enables a simpler description of a nonequilibrium process \cite{sek10}}.  Close to equilibrium, when $\rho_i= \rho_j + d\rho_{ij}$, the relative entropy $D(\rho_i|| \rho_j)$ can be   Taylor expanded as \cite{sal85,nul88},
\begin{equation}
D(\rho_i|| \rho_j) \simeq k_\mathrm{B}\int dzdv_z \frac{(d\rho_{ij})^2}{\rho_j} = k_\mathrm{B} \frac{Q_{ij}^2}{\sigma_{j}^2},
\end{equation}
where $\sigma_{j}^2 = \int dzdv_z (H - \langle H \rangle)^2 \rho_j$ is the  variance  measuring energy fluctuations. This  approximation, which expresses the  entropy production as the distance between the mean energies in units of the fluctuations, is shown (triangles) for both processes in Fig.~3. It is markedly different from the exact values (crosses), indicating that the two transformations operate far from equilibrium. 

In conclusion, we have presented a detailed  experimental study  of a reversed Clausius inequality using an isolated sample of a dilute gas of ultracold Cesium atoms. While it agrees with the usual Clausius inequality for equilibrium processes, it predicts exact opposite possible nonequilibrium transformations. It thus highlights the fact that the direction of a thermodynamic process is determined by the given initial conditions \cite{bol64,par08,mic19}. Like the  inequality derived by Clausius, it should be viewed as a mathematical expression of the second law of thermodynamics, in cases where an isolated equilibrium system is acted upon by a nonequilibrium heat source, and possibly an additional work source, $T\Delta S \leq \Delta U + W$,  in the absence of a heat bath. The reversed Clausius inequality not only applies  to  dilute cold-atom gases, but to all, classical and quantum, isolated systems that do not thermalize on their own, from nonlinear interacting systems, such as discrete breathers \cite{fla98} and Fermi-Pasta-Ulam-type systems \cite{for92},  to strongly coupled integrable many-body systems with long-lived transient states \cite{kin06,pol11}.

We acknowledge  financial support from the German Science Foundation (DFG) under Project No.~277625399 - TRR 185 and Grant No.~FOR 2724.

\section*{Supplemental Material}

\subsection{Laser cooling techniques}

For the optical molasses cooling of single Cs atoms, we employ the three-dimensional molasses laser setup which is also used for the creation of the magneto-optical trap~\cite{met99}.
There, three pairs of counter-propagating laser beams are employed for both, cooling and repumping light.
The total cooling light intensity is $I_\mathrm{cool} = \SI{7.3}{\milli\watt / \centi\meter\squared} = 2.7 I_\mathrm{sat}$ and the detuning of cooling light to the $F=4 \rightarrow F'=5$ hyperfine transition of the Cs D2 line~\cite{ste10} is varied between $\Delta f_\mathrm{cool} = \SI{-82}{\mega\hertz} = -15.7 \Gamma_\mathrm{D2}$ and $\Delta f_\mathrm{cool} = \SI{-42}{\mega\hertz} = -8.0 \Gamma_\mathrm{D2}$.
The repumping light at a total intensity of $I_\mathrm{rep} = \SI{2.5}{\milli\watt / \centi\meter\squared} = 0.9 I_\mathrm{sat}$ is on resonance to the $F=3 \rightarrow F'=4$ hyperfine transition.
When applying the optical molasses cooling to Cs atoms stored in the optical dipole trap, the atoms are cooled and at the same time move in the trapping potential.
This creates a thermal-phase space distribution, if the duration of the molasses pulse is sufficiently long compared to the inverse trap frequency.
For the molasses pulses employed here with several $\SI{100}{\milli\second}$ duration, the state created after the cooling pulse is hence a thermal Gibbs state.

This is in contrast to the state created by the degenerate Raman sideband cooling (DRSC) technique.
The DRSC scheme employed in our experiment closely follows Ref.~\cite{ker00}. An illustration of our setup can be found in {Ref.~\cite{may19}}.
A set of four DRSC lattice laser beams creates a three-dimensional optical lattice potential which is superposed to the dipole trap potential.
At a typical detuning of $\delta_{44} = \SI{-6}{\mega\hertz} = -1.1 \Gamma_\mathrm{D2}$ to the $\ket{F=4}\rightarrow\ket{F'=4}$-transition and an intensity of $\SI[per-mode=symbol]{0.48}{\watt\per\square\centi\meter}$, the trap depth of this optical lattice for atoms in the $\ket{F=4}$-manifold is around $k_\mathrm{B} \times \SI{40}{\micro\kelvin}$ with trap frequencies in the range of $10-\SI{100}{\kilo\hertz}$.
The cooling effect of the DRSC is based on the successive reduction of the vibrational state $\ket{n}$ in this lattice site.
This is achieved by adjusting the magnetic field such that neighboring levels $\ket{F, m_f}\ket{n}$ and $\ket{F, m_f+1}\ket{n-1}$ of hyperfine and vibrational state become degenerate and hence can be coupled by a Raman transition with two  photons from the DRSC lattice laser.
An additional DRSC pump beam driving $\sigma^+$-transitions at a detuning of $\delta_{32} = \SI{12}{\mega\hertz} = 2.3 \Gamma_\mathrm{D2}$ from the $\ket{F=3}\rightarrow\ket{F'=2}$-transition dissipates the vibrational energy.
As a result, for the duration of the DRSC pulse, the Cs atoms are confined in the lattice sites created by the DRSC lattice lasers and cannot move in the optical dipole trap.
Therefore, after the DRSC pulse, the atomic velocity distribution is given by the temperature $T_\text{R}$ which induced by the DRSC, but the position distribution in the optical dipole trap is unchanged and hence corresponds to a different temperature.

\subsection{Potential separation}

The anharmonic  trapping potential in the experiment can be well described by a separable potential, $\tilde{U}(x, y, z) \approx U_x(x) + U_y(y)+ U_z(z)$.
This approximation is constructed by sampling the full potential along the three principal axes $x$, $y$, and $z$, yielding the potential cuts $U_i(i)$.
Figure~\ref{fig:overview} shows the potential cuts $U_i(i)$ and quantifies the validity of this approximation along the $xyz$-diagonal, where deviations  are expected to be most significant.
The relative deviation $\Delta_\mathrm{rel} = [\tilde{U}(x, y, z) - U(x, y, z) ] /  U(x, y, z)$ between the separable approximation and the full potential is below $\SI{10}{\percent}$ for atoms with energies below $k_\mathrm{B} \times \SI{100}{\micro\kelvin}$.
This indicates that for typical atomic energies of $k_\mathrm{B} \times \SI{10}{\micro\kelvin}$ the separable approximation $\tilde{U}$ is not only qualitatively but also quantitatively a good approximation of the full potential $U$.
As a result, the atomic motion along the radial and axial directions are decoupled, thereby facilitating an independent treatment of the observed axial direction.

\bibliographystyle{apsrev4-1}

\end{document}

%% file: packages.tex
\usepackage[utf8x]{inputenc}
\usepackage{color}
\usepackage{bbm}

\usepackage{nicefrac}
\usepackage{amsfonts,amsmath,amssymb,stmaryrd}
\usepackage{braket}

 \usepackage[autostyle=true]{csquotes}

\usepackage{tabularx}
\usepackage{multirow} 
\usepackage{hhline}
\usepackage{graphicx}
\usepackage{subfigure}  
\usepackage{bbm} 
\usepackage{hyperref}
\usepackage[pdftex]{epsfig}
\usepackage{mathrsfs}
\usepackage{verbatim}
\usepackage{centernot}
\usepackage{ulem}
\usepackage{array}
\usepackage{cancel}
\usepackage{ifthen}
\usepackage{bm}	
\usepackage{siunitx}
\DeclareSIUnit\gauss{G}
\usepackage{todonotes}
\usepackage{setspace}
\usepackage{bm} 

\InputIfFileExists{gitinfo-latex.inc}{}{}

\usepackage{listings}
\usepackage{color}

\usepackage[acronym,nomain]{glossaries}

%% file: manuscript.bbl
\begin{thebibliography}{99}	
	\bibitem{cla79} R. Clausius, \textit{The Mechanical Theory of Heat}, (Macmillan, London, 1879).
	\bibitem{pip66} A. B. Pippard, {\it Elements of Classical Thermodynamics}, (Cambridge University Press, Cambridge, 1966).
	\bibitem{bus05} C. Bustamante, J. Liphardt, and  F. Ritort, The nonequilibrium thermodynamics of small systems, Physics Today \textbf{58},
	43 (2005).
	\bibitem{sek10}   K. Sekimoto, \textit{Stochastic Energetics}, (Springer, Berlin, 2010).
	\bibitem{sei12}U. Seifert, Stochastic thermodynamics, fluctuation theorems, and molecular machines, Rep. Prog. Phys. \textbf{75}, 126001 (2012).
	\bibitem{jar11} C. Jarzynski, Equalities and Inequalities: Irreversibility and the Second Law of Thermodynamics at the Nanoscale, Annu. Rev. Condens. Matter Phys. \textbf{2}, 329 (2011).
	\bibitem{cil13}S. Ciliberto, R. Gomez-Solano, and A. Petrosyan, Fluctuations, Linear Response, and Currents in Out-of-Equilibrium Systems, Annu. Rev. Condens. Matter Phys. \textbf{4}, 235 (2013).
	\bibitem{nie00} M. A. Nielsen and I. L. Chuang, \textit{Quantum Computation and Quantum Information}, (Cambridge University Press, Cambridge, 2000).
	\bibitem{dow03}  J. P. Dowling and G. J. Milburn, Quantum Technology: The Second Quantum Revolution, Phil. Trans. R. Soc. A \textbf{361}, 3655 (2003).
	\bibitem{eis15} J. Eisert, M. Friesdorf, and C. Gogolin, Quantum many-body systems out of equilibrium, Nature Phys. \textbf{11}, 124 (2015).
	\bibitem{gog16} C. Gogolin and J. Eisert, Equilibration, thermalisation, and the emergence of statistical mechanics in closed quantum systems, Rep. Prog.  Phys. 79, 056001 (2016).
	\bibitem{par89} H. Partovi, Quantum thermodynamics, Phys. Lett. A \textbf{137}, 440 (1989).
	\bibitem{per02} A. Peres, \textit{Quantum Theory: Concepts and Methods}, (Kluwer Academic Publishers, New York, 2002), Chap.~9.

	\bibitem{gav16} B. Gaveau, L. Granger, M. Moreau, and L. S. Schulman, Relative Entropy, Interaction Energy and the Nature of Dissipation, Entropy \textbf{16}, 3173 (2016).
	\bibitem{met99} H. J. Metcalf and P. van der Straten, \textit{Laser Cooling and Trapping}, (Springer, Berlin, 1999). 
	\bibitem{coh11} C. Cohen-Tannoudji and D. Guery-Odelin, \textit{Advances in Atomic Physics}, (World Scientific, Singapore, 2011).
	
	\bibitem{vul98} V. Vuletic, C. Chin, A. J. Kerman, and S. Chu, Degenerate Raman Sideband Cooling of Trapped Cesium Atoms at Very High Atomic Densities, Phys. Rev. Lett. \textbf{81}, 5768 (1998).
	\bibitem{ker00} A. J. Kerman, V. Vuletic, C. Chin, and S. Chu, Beyond Optical Molasses: 3D Raman Sideband Cooling of Atomic Cesium to High Phase-Space Density, Phys. Rev. Lett. \textbf{84}, 439 (2000).
	\bibitem{han00} D.-J. Han, S. Wolf, S. Oliver, C. McCormick, M. T. DePue, and D. S. Weiss, 3D Raman Sideband Cooling of Cesium Atoms at High Density, Phys. Rev. Lett. \textbf{85}, 724 (2000).
	
	\bibitem{sch16} F. Schmidt, D. Mayer, M. Hohmann, T. Lausch, F. Kindermann, and A. Widera, Precision measurement of the $^{87}\text{Rb}$ tune-out wavelength in the hyperfine ground state $F=1$ at 790 nm, Phys. Rev. A \textbf{93}, 022507 (2016).
	
\bibitem{pro76} I. Procaccia and R. D. Levine, Potential work: A statistical-mechanical approach for systems in disequilibrium,J. Chem. Phys. \textbf{65}, 3357 (1976).
\bibitem{sch80} F. Schl\"ogl, Stochastic Measures in nonequilibrium thermodynamics, Phys. Rep. \textbf{62}, 287 (1980).
\bibitem{esp10} M. Esposito, K. Lindenberg, and C. V. d. Broeck, Entropy production as correlation between system and reservoir, New J. Phys. \textbf{12}, 013013 (2010).
\bibitem{def11} S. Deffner and E. Lutz, Nonequilibrium entropy production for open quantum systems, Phys. Rev. Lett. \textbf{107}, 140404 (2011).
\bibitem{cov06} T. M. Cover and J. A. Thomas, \textit{Elements of Information Theory}, (Wiley, New York, 2006).
\bibitem{jay57}  E. T. Jaynes, Information Theory and Statistical Mechanics, Phys. Rev. \textbf{106}, 620 (1957).
\bibitem{bol64} L. Boltzmann, \textit{Lectures on Gas Theorie}, (Dover, New York, 1964).
\bibitem{ste10} D. A. Steck, Cesium D Line Data \url{https://steck.us/alkalidata/cesiumnumbers.pdf} (2010).
\bibitem{mil86} G. J. Milburn, Quantum and classical Liouville dynamics of the anharmonic oscillator, Phys. Rev. A \textbf{33}, 674 (1986).
\bibitem{sal85} P. Salamon, J. D. Nulton, and R. S. Berry, Length in statistical thermodynamics,  J. Chem. Phys. \textbf{82}, 2433 (1985).
\bibitem{nul88}  J. D. Nulton and P. Salamon, Statistical mechanics of combinatorial optimization, Phys. Rev. A \textbf{37}, 1351 (1988).
\bibitem{par08} Partovi, M. H., Entanglement versus stosszahlansatz:
disappearance of the thermodynamic arrow in a high-correlation environment.
{Phys. Rev. E} \textbf{77}, 021110 (2008).
\bibitem{mic19} K. Micadei, J. P. S. Peterson, A. M. Souza, R. S. Sarthour, I. S. Oliveira, G. T. Landi, T. B. Batalh\~ao, R. M. Serra, and E. Lutz, Reversing the direction of heat flow using quantum correlations, Nature Comm. {\bf 10}, 2456 (2019).	

\bibitem{fla98} S. Flach, C. R. Willis, Discrete breathers, Phys. Rep. \textbf{295} 181 (1998).
\bibitem{for92} J. Ford, The Fermi-Pasta-Ulam problem: Paradox turns discovery, Phys. Rep. \textbf{21}3, 271 (1992).
\bibitem{kin06} T. Kinoshita, T. Wenger, and D. S. Weiss, A quantum Newton's cradle, Nature \textbf{440}, 900 (2006).
\bibitem{pol11} A. Polkovnikov, K. Sengupta, A. Silva, and M. Vengalattore, Colloquium: Nonequilibrium dynamics of closed interacting quantum systems, Rev. Mod. Phys. \textbf{83}, 863 (2011).
\bibitem{may19} D. Mayer, F. Schmidt, D. Adam, S. Haupt, J. Koch, T. Lausch, Jens Nettersheim, Q. Bouton, and A. Widera, J. Phys. B: At. Mol. Opt. Phys. \textbf{52}, 015301 (2019).
	
\end{thebibliography}
